\documentclass[english]{article}
\usepackage{lmodern}

\usepackage[T1]{fontenc}
\usepackage[latin9]{inputenc}
\usepackage{amsbsy}
\usepackage{amstext}
\usepackage{graphicx}

\makeatletter
\newcommand{\lyxaddress}[1]{
\par {\raggedright #1
\vspace{1.4em}
\noindent\par}
}

\usepackage{mathrsfs}   
\usepackage{slashed}     
\usepackage{bbold}  
\usepackage{url}
\usepackage{graphicx}
\usepackage[colorlinks=true,linkcolor=redLinks,citecolor=greenLinks,urlcolor=redLinks, pdfborder={0 0 1}]{hyperref}
\usepackage{xcolor}
\usepackage{framed}
\usepackage[numbers,sort&compress]{natbib}
\usepackage{amsmath}
\usepackage[frozencache]{minted}

\allowdisplaybreaks

\colorlet{shadecolor}{gray!15}

\definecolor{greenLinks}{rgb}{0, 0.6, 0} 
\definecolor{blueLinks}{rgb}{0, 0, 0.6}
\definecolor{redLinks}{rgb}{0.6, 0, 0}
\definecolor{tempText}{rgb}{0.55, 0.10,0.67}
\definecolor{eprintLinks}{rgb}{0.4, 0.4, 0.4}
\definecolor{journalLinks}{rgb}{0.6, 0, 0}

\newcommand{\MYhref}[3][redLinks]{\href{#2}{\color{#1}{#3}}}%

\usepackage{multirow}
\let\orig@Hy@EveryPageAnchor\Hy@EveryPageAnchor
\def\Hy@EveryPageAnchor{%
    \begingroup
    \hypersetup{pdfview=Fit}%
    \orig@Hy@EveryPageAnchor
    \endgroup
}

\let\oldFootnote\footnote
\newcommand\nextToken\relax

\renewcommand\footnote[1]{%
    \oldFootnote{#1}\futurelet\nextToken\isFootnote}

\newcommand\isFootnote{%
    \ifx\footnote\nextToken\textsuperscript{,}\fi}

\makeatother

\usepackage{babel}
\begin{document}

\title{The \texttt{Sym2Int} program: going from symmetries to interactions}

\author{Renato M. Fonseca\thanks{E-mail: renato.fonseca@ific.uv.es} \thanks{Contribution to the proceedings of the Fifth Symposium on Prospects in the Physics of Discrete Symmetries (DISCRETE 2016), Warsaw, Poland, 28 November -- 3 December 2016.} 
\date{}}

\maketitle

\lyxaddress{\begin{center}
{\Large{}\vspace{-0.5cm}}AHEP Group, Instituto de Física Corpuscular,
C.S.I.C./Universitat de València\\
Parque Científico, Calle Catedrático José Beltrán, 2 E-46980 Paterna
(Valencia) -- Spain
\par\end{center}}
\begin{abstract}
Model builders often need to find the most general Lagrangian which
can be constructed from a given list of fields. These fields are actually
representations of the Lorentz and gauge groups (and maybe of some
discrete symmetry group as well). I will describe a simple program
(\texttt{Sym2Int}) which helps to automate this task by listing all
possible interactions between Lorentz/gauge group representations.
\end{abstract}

\section{Symmetries of the fields and their interactions}

Symmetries are an ever present feature in beyond-the-Standard-Model
models: there is the Lorentz symmetry of flat space-time, there is
the gauge symmetry containing $SU(3)_{c}\times SU(2)_{L}\times U(1)_{Y}$,
and perhaps there are even more symmetries (for example discrete ones
associated to flavor). Therefore, to build a model one first needs
to find the most general Lagrangian which can be formed with a given
list of fields, each transforming under some irreducible representation
of the full symmetry group.

This is very often (or perhaps always) done manually, taking time
and potentially leading either to the accidental inclusion of forbidden
interactions, or the omission of some of the allowed ones. Furthermore,
it is worth emphasizing the following two points. The first one is
that the list of interactions in the Standard Model (SM) is small,
but with the addition of just a few extra fields their number can
increase considerably. The second thing worth keeping in mind is that
all SM fields transform under small representations of the gauge group,
which simplifies the task of manually computing the allowed interactions.

The aim of the \texttt{Sym2Int} package for Mathematica is to automatize
this task: with minimal effort one can quickly obtain a full picture
of a model's interactions and at the same time reduce the probability
of making mistakes. It also introduces the possibility of cross-checking
results obtained by hand. The program can be downloaded from the webpage

\begin{center}
\href{http://renatofonseca.net/sym2int.php}{renatofonseca.net/sym2int.php}
\par\end{center}

\begin{figure}
\begin{centering}
\includegraphics[scale=0.82]{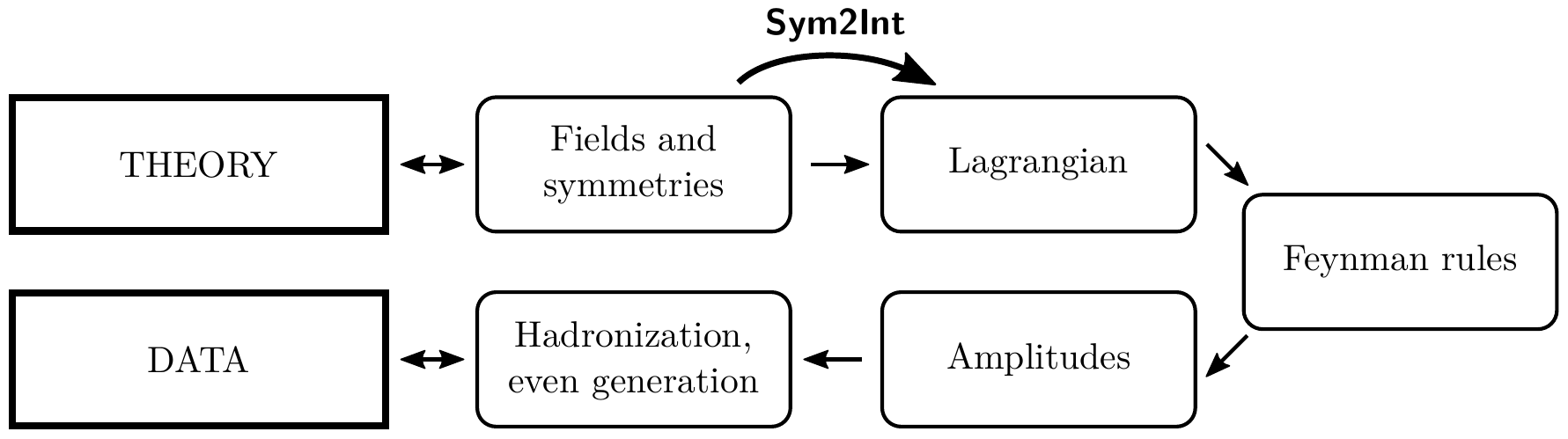}
\par\end{centering}

\protect\caption{\label{fig:ChainOfSteps}Chain of steps needed in order to explore
the phenomenology of a model. The \texttt{Sym2Int} package helps with
the first one, by automatizing the task of finding the interactions
allowed by symmetries.}
\end{figure}

There are already many tools which automate various of the steps required
to confronted theory with data (see figure \ref{fig:ChainOfSteps}).
However, to my knowledge the very first step described previously
(i.e, finding a model's interactions) has not been addressed fully.
In fact, with the notable exception of \texttt{Susyno} \cite{Fonseca:2011sy}
and \texttt{SARAH} \cite{Staub:2013tta}, most tools assume as a starting
point the Lagrangian (at the very least). In this sense, the \texttt{Sym2Int}
program (which uses \texttt{Susyno} functions) can be seen as a missing
link in this long chain of programs which are needed to automatize
the study of the phenomenology of a model. As we will see latter on,
the code counts and lists the allowed interactions of a model, but
it can also provide the explicit contractions of the field components
(for example, the color indices, the $SU(2)_{L}$ indices, and so
on) which are needed to write down the Lagrangian. 

Going from fields and symmetries to a Lagrangian requires group theory
code, which is available in \texttt{Susyno}. In fact, this program
already takes as input symmetries and fields/representations, building
from there a Lagrangian. However, it does so for supersymmetric models
and furthermore there is an emphasis on the calculation of renormalization
group equations. The \texttt{Sym2Int} package uses extensively the
group theory code of \texttt{Susyno} but gets rid of the emphasis
on theses two aspects (supersymmetry and renormalization group equations):
instead, for a generic model, \texttt{Sym2Int} simply focuses on providing
a list of allowed couplings and, if the user so wishes, it also shows
how to contract the indices of the fields which participate in each
interaction.

The program can deal with any representation of the Lorentz and gauge
groups, and it will also work with abelian discrete symmetries. There
is currently no support for non-abelian discrete symmetries; this
will hopefully be added in the future. As for the type of interactions,
\texttt{Sym2Int} will compute all allowed terms up to an arbitrary
mass dimension specified by the user (it is therefore not limited
to renormalizable operators). Unfortunately, due to their peculiar
transformation property under gauge transformations, non-renormalizable
terms with derivatives $\partial_{\mu}$ and/or gauge bosons are not
yet handled by the program in a fully satisfactory way.\footnote{This shortcoming will probably also be addressed in a latter version
of the code. Regarding this matter, I would like to thank Andreas
Trautner for bringing to my attention reference \cite{Henning:2015daa}.}

Having introduced the \texttt{Sym2Int} program, in the rest of this
document I will describe how to use it. This information and some
more details are available on the program's webpage \cite{Sym2Int_Web};
furthermore, the webpage \cite{Susyno_GT_Web} may be of some use
since it explains how to use \texttt{Susyno}'s group theory code in
a stand-alone way.

\section{Installing and running the program}

The \textit{sym2Int.m} file which can be obtained from the program's
webpage \cite{Sym2Int_Web} should be placed in a folder where Mathematica
can find it. This could be for example \textit{(Mathematica base dir.)/AddOns/Applications}
on Linux and Mac OS or \textit{(Mathematica base dir.)\textbackslash{}AddOns\textbackslash{}Applications}
on Windows. Since the program relies on \texttt{Susyno} code, this
program must also be installed (it is available from the webpage \cite{Susyno_Web}).

Once this is done, \texttt{Sym2Int} can be loaded with the command

\definecolor{lightGray}{rgb}{0.95,0.95,0.95}
\begin{minted}[escapeinside=||,bgcolor=lightGray]{c}
<<Sym2Int|`|
\end{minted}

\section{\label{sec:3}Defining a model}

The program assumes that the defining elements of a model are (i)
a \textbf{gauge group} and (ii) a \textbf{list of fields} which are
irreducible representations of the gauge and Lorentz groups. For bookkeeping
reasons, a name for the model is also required:

\begin{minted}[mathescape,escapeinside=||,bgcolor=lightGray]{mathematica}
gaugeGroup[<model name>] ^= <gauge group>;
fields[<model name>] ^= {<field 1>, <field 2>, |...|};
GenerateListOfCouplings[<model name>];
\end{minted}

In turn, each field consists of (i) a \textbf{name}, (ii) a \textbf{gauge
representation}, (iii) a \textbf{Lorentz representation} {[}such as
scalar (\texttt{``S''}), right/left-handed Weyl fermion (\texttt{``R''}/\texttt{``L''}),
vector (\texttt{``V''}), etc.{]}, (iv) an indication of whether
the \textbf{field is real} (\texttt{\textquotedbl{}R\textquotedbl{}})
\textbf{or complex} (\texttt{\textquotedbl{}C\textquotedbl{}}), and
(v) the \textbf{number of copies/flavors}.

\begin{minted}[mathescape,escapeinside=||,bgcolor=lightGray]{mathematica}
<field> = {<name>, <gauge rep>, <Lorentz rep|:| "S", "R", "L", "V", etc|.|>, <"R" or "C">,
           <|#|copies>}
\end{minted}

For example, the SM can be specified as follows:

\begin{minted}[mathescape,escapeinside=||,bgcolor=lightGray]{mathematica}
gaugeGroup[SM] ^= {SU3, SU2, U1};

fld1 = {"u", {3, 1, 2/3}, "R", "C", 3};
fld2 = {"d", {3, 1, -1/3}, "R", "C", 3};
fld3 = {"Q", {3, 2, 1/6}, "L", "C", 3};
fld4 = {"e", {1, 1, -1}, "R", "C", 3};
fld5 = {"L", {1, 2, -1/2}, "L", "C", 3};
fld6 = {"H", {1, 2, 1/2}, "S", "C", 1};
fields[SM] ^= {fld1, fld2, fld3, fld4, fld5, fld6};

GenerateListOfCouplings[SM];
\end{minted}

With this input, the program prints the following table with a list
of all possible renormalizable interactions among the given fields
(table \ref{tab:SM}).

\begin{center}
\begin{table}[tbph]
\begin{centering}
\includegraphics[scale=0.82]{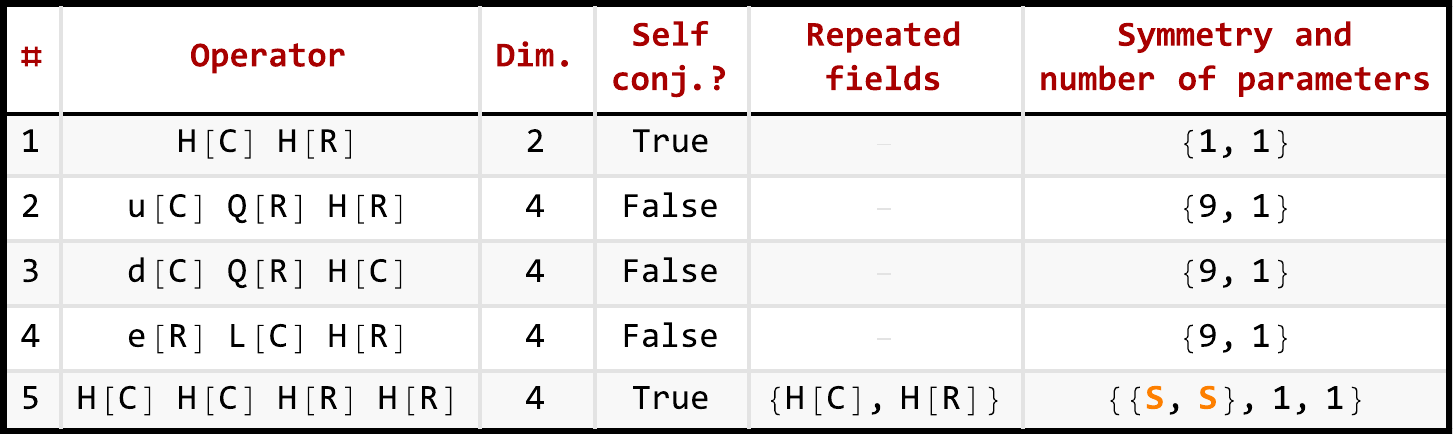}
\par\end{centering}

\protect\caption{\label{tab:SM}Standard Model renormalizable interactions (those with
gauge bosons are not shown).}
\end{table}

\par\end{center}

By default, the list stops at mass dimension 4. To go further, for
example up to dimension 5, one can use the \texttt{MaxOrder} command:

\begin{minted}[mathescape,escapeinside=||,bgcolor=lightGray]{mathematica}
GenerateListOfCouplings[SM, MaxOrder -> 5];
\end{minted}

The output is the one shown in table \ref{fig:SMdim5}, which contains
an extra row corresponding to the Weinberg operator.
\begin{table}[tbph]
\begin{centering}
\includegraphics[scale=0.82]{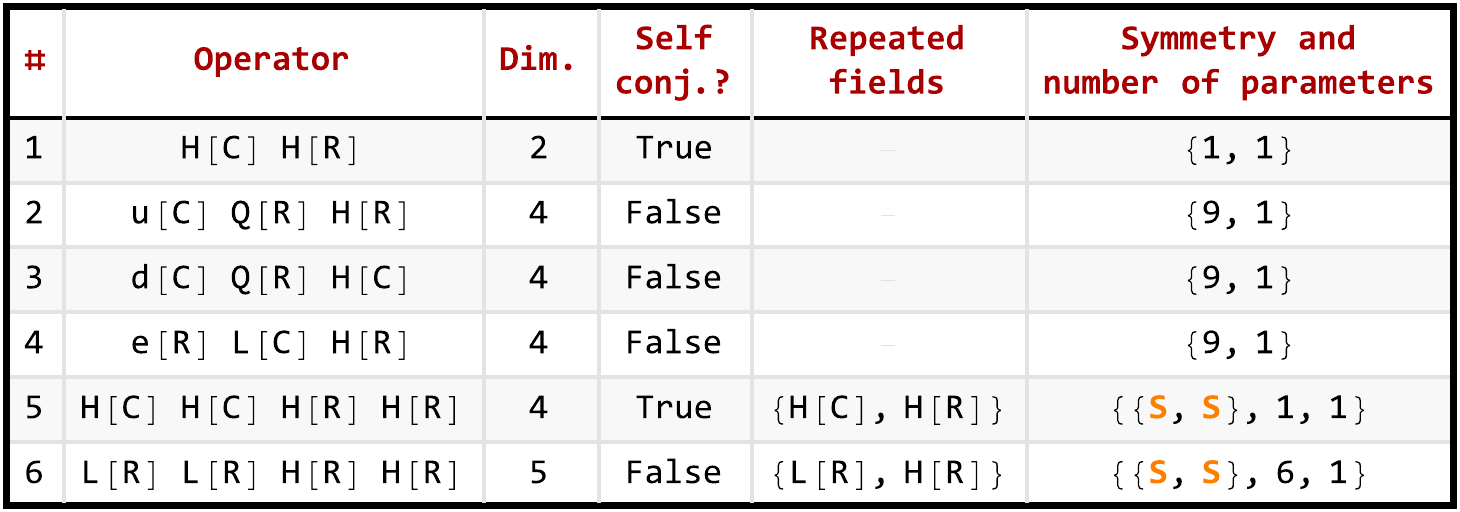}
\par\end{centering}

\protect\caption{\label{fig:SMdim5}Standard Model interactions up to dimension 5 (those
with gauge bosons or derivatives are not shown).}
\end{table}

Let us now take a look at the information contained in these tables
which are printed by the program. For that discussion, it will be
convenient to use the word \textit{operator} in a rather liberal way:
it stands for all gauge and Lorentz invariant contractions of a given
combination of fields. As such, even if there is more than one independent
gauge and Lorentz invariant contraction of the fields (say) A, B,
C and D, I will still consider them to be all part of a single ABCD
operator. So, with this clarification, the output tables contains
the following information:
\begin{itemize}
\item The \textbf{first column} simply numbers each operator.
\item The \textbf{second column} indicates what are the fields which participate
in the interaction; a \textquotedbl{}\texttt{{[}C{]}}\textquotedbl{}
after the field's name means that it is conjugated, while a \textquotedbl{}\texttt{{[}R{]}}\textquotedbl{}
means that it is not.
\item The \textbf{third column} shows, for convenience, the mass dimension
of the operator: each boson(fermion) contributes $1$($3/2$) unit(s).
\item The \textbf{forth column} tells us whether or nor the operator is
self-conjugated. 
\item The \textbf{fifth column} lists the fields which appear more than
once in the operator (taking into account that conjugated fields with
a ``\texttt{{[}C{]}}'' are not the same as the unconjugated ones
with a ``\texttt{{[}R{]}}'').
\item The \textbf{sixth column} contains information on the permutation
symmetry and number of parameters associated to the operator. In general
is has the structure ``\texttt{\{<symmetry (if any)>, <nFlav>, <nContr>\}}''.

\begin{itemize}
\item If there is any repeated field(s) in the operator, the \textbf{first
entry} indicates what is the permutation symmetry associated to them.
Otherwise, this entry is omitted.
\item Some fields have flavors, i.e. they come with a multiplicity. So,
the \textbf{next entry} indicates the number of couplings/numbers
needed to encode each independent contraction of the gauge and Lorentz
indices of the fields.
\item The \textbf{last entry} indicates the number of independent contractions
of the gauge and Lorentz indices of the fields.\footnote{This is true in most cases. However, in cases with complex permutation
symmetries, this description ceases to be accurate. For the correct/full
description of this number, see the discussion in section \ref{sec:4}
concerning table \ref{fig:modelX}.}
\end{itemize}
\end{itemize}
All this should be clear, with perhaps the exception of the last column,
which deals with parameter counting and also with symmetries under
permutations of fields. This is a subtle yet very important issue
which we will now discuss in detail.

\section{\label{sec:4}Counting parameters}

The number of parameters needed to encode a model is not the same
as the number of rows in the output table for various simple reasons.
Let us go through them:
\begin{enumerate}
\item The number of copies of each field must be taken into account. For
example, in the above SM example, there is a $u^{*}QH$ interaction\footnote{The star represents just field conjugation. I omit the exact contraction
of Lorentz and gauge indices. } on row \#2, but since both $u$ and $Q$ have 3 flavors we need a
tensor $Y_{ij}$ with dimensions $3\times3$ to encode all the couplings
of this operator. There is then a need for 9 parameters, and the program
presents this counting on the the second(first) entry of the last
column is there are(there aren't) repeated fields in the operator.
\item The coupling multiplying a combination of fields $\mathcal{O}$ is
real if $\mathcal{O}=\mathcal{O}^{*}$ and complex otherwise, meaning
that operators which are self-conjugate (forth column of the output
matrix) require half as many real couplings as those which are not
self-conjugate.\footnote{Once we introduce flavors things may become a bit more complicated.
For example, in a self-conjugate term $X_{ij}\phi_{i}^{*}\phi_{j}$
where the indices $i,j=1,\cdots,n$ stand for different flavors of
a field $\phi$, it is not true that $X_{ij}$ equals $X_{ij}^{*}$.
Instead, on must make a permutation of the $\left(ij\right)$ indices
jointly with the conjugation operation: $X_{ij}=X_{ji}^{*}$. However,
note that it still holds that $X_{ij}$ contains $n^{2}$ real degrees
of freedom (and not $2n^{2}$ real $=n^{2}$ complex degrees of freedom).} For example, the 9 parameters needed for the $u^{*}QH$ operator
in the SM are complex, while $H^{*}H$ requires a single real parameter.
\item Sometimes it is possible to contract the gauge and Lorentz indices
of the fields in more than one way: for example, there are two terms
of the form $\Delta^{*}\Delta^{*}\Delta\Delta$ for an $SU(2)$ complex
scalar triplet $\Delta$. This number is shown in the last entry of
the sixth column.
\item Finally, consider the well-known non-renormalizable operator $L_{i}L_{j}HH$
responsible for neutrino mass generation in the SM (row \#6 of table
\ref{fig:SMdim5}). The flavor indices $i$ and $j$ go from 1 to
3, so a $3\times3$ coupling tensor $\kappa_{ij}$ is required. Therefore,
naively one would think from the previous considerations that there
are $9$ complex degrees of freedom association to this operator.
However, it is well known that the leptons $L$ contract symmetrically
in this particular case, hence $\kappa_{ij}=\kappa_{ji}$, so there
are only 6 complex parameters and indeed that is the number displayed
on the last column (second entry). In fact, the two Higgs fields also
contract symmetrically (otherwise, since there is only one copy of
this field, there would be no operator!). That is what the \texttt{``\{``S'',
``S''\}''} in the first entry of the last column means: there are
two fields which appear more than once in the operator (``\texttt{L{[}R}{]}''
and ``\texttt{H{[}R{]}}'' as indicated in the fifth column) and
both of them contract symmetrically (an ``\texttt{A}'' would indicate
an anti-symmetry).
\end{enumerate}
In practice, this discussion implies that the number of couplings
associated to each operator is obtained by making the product of the
two numbers in the sixth column. However, note that sometimes an operator
might have different contractions of the gauge/Lorentz components
of the fields with different symmetries; in that case the sixth column
will be of the form ``\texttt{\{<symmetry1>, <nFlav1>, <nContr1>\}
| \{<symmetry2>, <nFlav2>, <nContr2>\} | ...}'' and the number of
degrees of freedom is given by the sum of the products \texttt{<nFlavI><nContrI>}.

This last point is better illustrated with another example: the Two
Higgs Doublet Model (THDM). There are two ways of inserting a second
Higgs in the SM: one of them is to say that the fields are still ``\texttt{\{fld1,
fld2, fld3, fld4, fld5, fld6\}}'' but now there are two copies of
the Higgs:

\begin{minted}[mathescape,escapeinside=||,bgcolor=lightGray]{mathematica}
fld6 = {"Hi", {1, 2, 1/2}, "S", "C", 2};
\end{minted}

Another possibility is to keep ``\texttt{fld6}'' as it was before,
and add a ``\texttt{fld7}'' with the same quantum numbers:

\begin{minted}[mathescape,escapeinside=||,bgcolor=lightGray]{mathematica}
fld6 = {"H1", {1, 2, 1/2}, "S", "C", 1};
fld7 = {"H2", {1, 2, 1/2}, "S", "C", 1};
fields[THDM] ^= {fld1, fld2, fld3, fld4, fld5, fld6, fld7};
\end{minted}

In the first case, we get table \ref{fig:2HDM_v1}, while in the second
one we get table \ref{fig:2HDM_v2}. At first sight, these two results
seem very different, but it is easy to check that they represent the
same list of interactions. Table \ref{fig:2HDM_v1} is more compact
because the two scalar doublets are treated as two copies/flavours
of a single field ``\texttt{H}i''; for example, the quartic couplings
are all encoded in row \#5 of table \ref{fig:2HDM_v1} while in table
\ref{fig:2HDM_v2} they are shown on the last six rows.\footnote{The counting of quartic parameters in the two cases goes as follows.
The operator associated to row \#5 of table \ref{fig:2HDM_v1} requires
$9\times1+1\times1=10$ real couplings. As for table \ref{fig:2HDM_v2},
rows \#10, \#13 and \#15 correspond to self-conjugated operators,
so they need a total of $1\times1+1\times2+1\times1=4$ real couplings,
while rows \#11, \#12 and \#14 need a total of $1\times1+1\times1+1\times1=3$
complex couplings, hence there is a total of 10 degrees of freedom
in the quartic sector --- just as in table \ref{fig:2HDM_v1}.}
\begin{table}[tbh]
\begin{centering}
\includegraphics[scale=0.82]{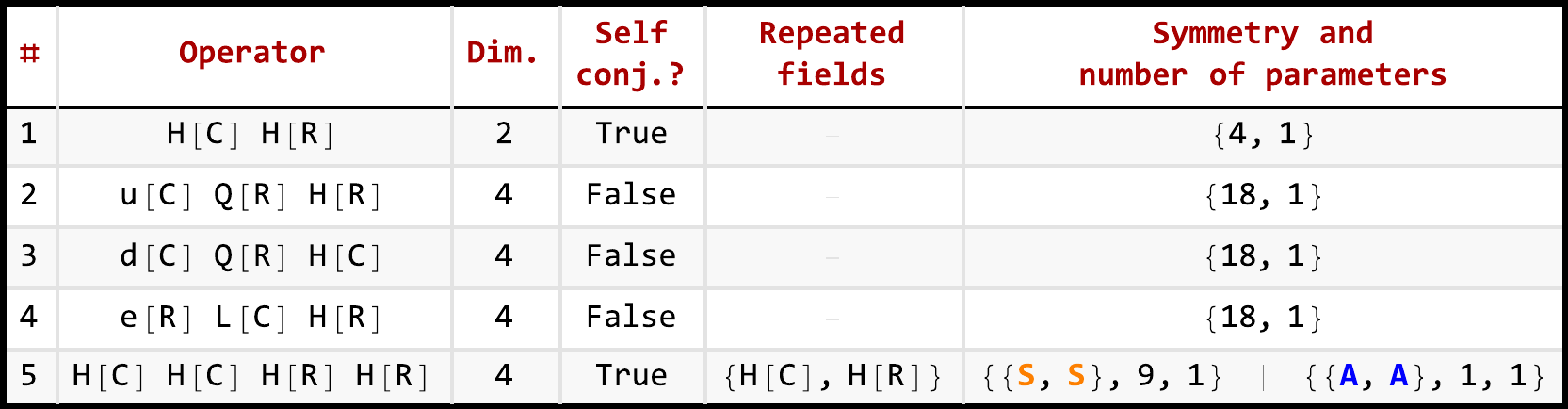}
\par\end{centering}

\protect\caption{\label{fig:2HDM_v1}List of interactions in the Two Higgs Doublet
Model (those with gauge bosons are not shown).}
\end{table}

\begin{center}
\begin{table}[tbh]
\begin{centering}
\includegraphics[scale=0.82]{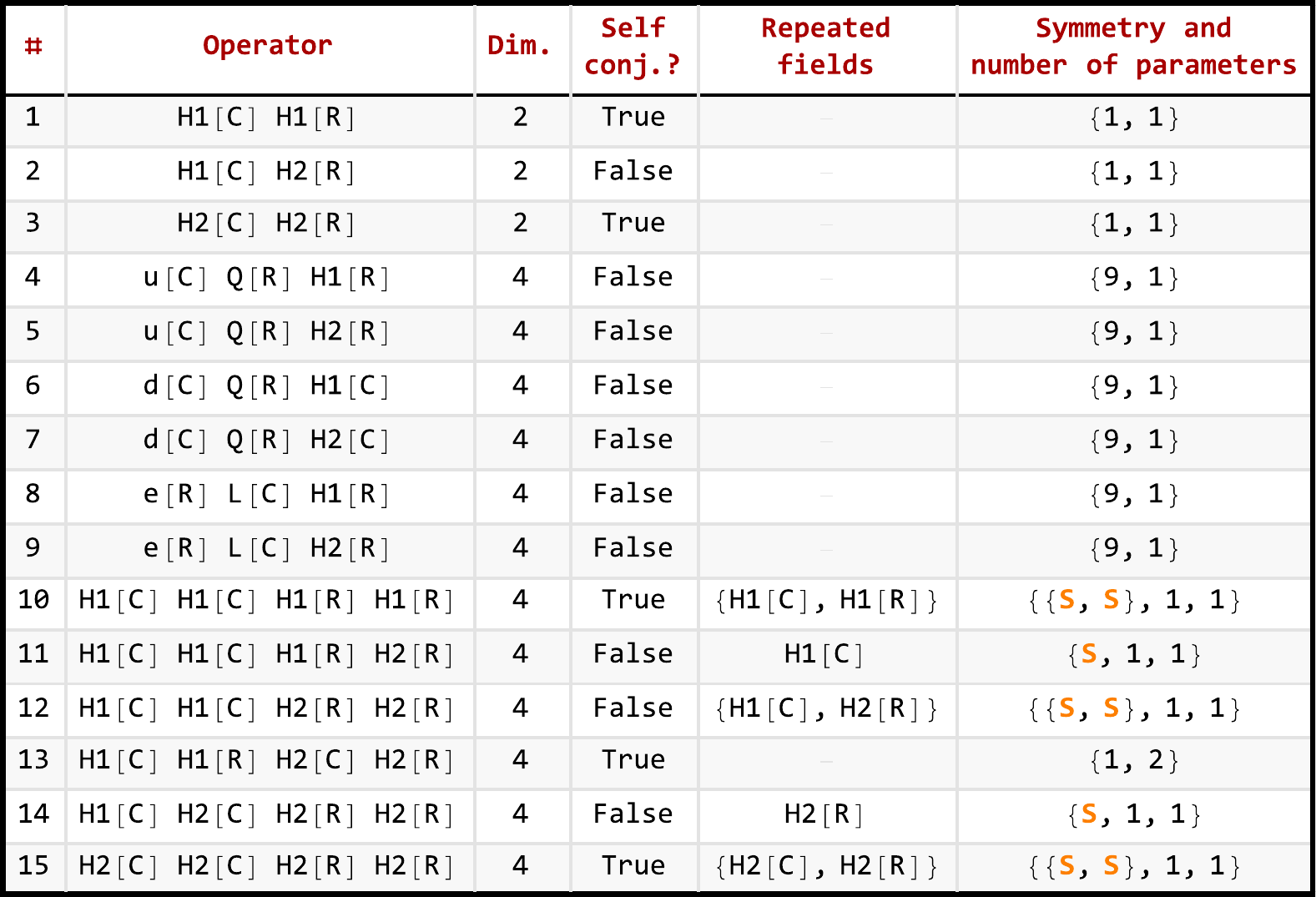}
\par\end{centering}

\protect\caption{\label{fig:2HDM_v2}List of interactions in the Two Higgs Doublet
Model (those with gauge bosons are not shown). This result may appear
different from the one in table \ref{fig:2HDM_v1}, but in reality
it is the same: both tables list the same interactions.}
\end{table}

\par\end{center}

Let us now focus on the first perspective, namely that there is a
single scalar field with two flavors: $H_{i}$ with $i=1,2$. Row
\#5 of table \ref{fig:2HDM_v1} is saying that one can write the quartic
interactions as
\begin{equation}
V^{(4)}=\lambda_{ijkl}^{SS}\left(H_{i}^{*}H_{j}^{*}H_{k}H_{l}\right)_{SS}+\lambda_{ijkl}^{AA}\left(H_{i}^{*}H_{j}^{*}H_{k}H_{l}\right)_{AA}\,,\label{eq:V4}
\end{equation}
where $\left(\cdots\right)_{SS}$ and $\left(\cdots\right)_{AA}$
stand for two different ways of contracting the $SU(2)_{L}$ indices
of the doublets: one is symmetric under the exchange of both the $\left(ij\right)$
and $\left(kl\right)$ indices, while the other contraction is anti-symmetric
under both these exchanges.\footnote{\label{fn:The-explicit-form}The explicit form of these contractions
is quite simple: $\left(H_{i}^{*}H_{j}^{*}H_{k}H_{l}\right)_{SS}=\left(H_{i}^{\dagger}H_{k}\right)\left(H_{j}^{\dagger}H_{l}\right)+\left(H_{j}^{\dagger}H_{k}\right)\left(H_{i}^{\dagger}H_{l}\right)$
and $\left(H_{i}^{*}H_{j}^{*}H_{k}H_{l}\right)_{AA}=\left(H_{i}^{\dagger}H_{k}\right)\left(H_{j}^{\dagger}H_{l}\right)-\left(H_{j}^{\dagger}H_{k}\right)\left(H_{i}^{\dagger}H_{l}\right)$
(up to some arbitrary normalization factors). Still, we will not be
discussing explicit contraction of indices since that is not necessary
for counting parameters/degrees of freedom of a model.} These symmetries are transmitted to the tensors of couplings $\lambda_{ijkl}^{SS}$
and $\lambda_{ijkl}^{AA}$, so that $\lambda_{ijkl}^{SS}=\lambda_{jikl}^{SS}=\lambda_{ijlk}^{SS}$
while $\lambda_{ijkl}^{AA}=-\lambda_{jikl}^{AA}=-\lambda_{ijlk}^{AA}$.
It is then straightforward to show that $\lambda_{ijkl}^{SS}$ is
described by 9 independent real numbers, while $\lambda_{ijkl}^{AA}$
requires a single one.

The origin of the two terms in equation (\ref{eq:V4}) can be traced
back to the fact that the product $\overline{\mathbf{2}}\times\overline{\mathbf{2}}\times\mathbf{2}\times\mathbf{2}$
in $SU(2)$ contains two singlets. But, even so, it is worth mentioning
that one does not need to split $V^{(4)}$ in two terms as in equation
(\ref{eq:V4}). For example, consider the sum $\left(\cdots\right)_{SS}+\left(\cdots\right)_{AA}\equiv\left(\cdots\right)_{mixed}$.
Then one could just write
\begin{equation}
V^{(4)}=\lambda_{ijkl}^{mixed}\left(H_{i}^{*}H_{j}^{*}H_{k}H_{l}\right)_{mixed}\label{eq:V4-2}
\end{equation}
since this is the same as $\lambda_{\left(ij\right)\left(kl\right)}^{mixed}\left(H_{i}^{*}H_{j}^{*}H_{k}H_{l}\right)_{SS}+\lambda_{\left[ij\right]\left[kl\right]}^{mixed}\left(H_{i}^{*}H_{j}^{*}H_{k}H_{l}\right)_{AA}$
where round(square) brackets in the indices denotes symmetrization(anti-symmetrization).
So there is a trade-off: the notation gets more compact if we replace
equation (\ref{eq:V4}) by equation (\ref{eq:V4-2}) but, in the latter
case, the counting of independent parameters is not as obvious. This
observation also highlights the fact that, if we hide different field
copies in a flavor index, then there might very well be an ambiguity
in the counting of interaction terms (as in equations (\ref{eq:V4})
and (\ref{eq:V4-2}), which actually stand for the same quartic potential).
The approach of the \texttt{Sym2Int} program is to always split the
interaction terms according to their properties under field permutation
symmetries, which makes the counting of parameters easier.

~

Field permutation symmetries can be more complicated than what has
been seen so far, and for completeness we should look at a more general
case. Consider the quartic operator of table \ref{fig:2HDM_v1}: both
$H_{i}^{*}$ and $H_{i}$ appear twice, hence the relevant permutation
group is $S_{2}\times S_{2}$. Crucially, the $S_{2}$ group only
has two 1-dimensional (irreducible) representations: the trivial/symmetric
(S) and the alternating/anti-symmetric (A) one. What happens if there
are more than $n>2$ repeated fields participating in an interaction?
Take, for example, a model with $m$ real scalar doublets $D_{i=1,\cdots,m}$
with no hypercharge.\footnote{This would lead to fractional electric charges, but we do not care
about such things here.} Is it possible to write down quartic interactions of the form $D_{i}D_{j}D_{k}D_{l}$?
If yes, how many couplings are needed as a function of $m$? To answer
these questions one needs to understand how do four $SU(2)$ doublets
contract. It is well known that $\mathbf{2}\times\mathbf{2}=\mathbf{1}_{A}+\mathbf{3}_{S}$,
so it is tempting to think that $\mathbf{2}\times\mathbf{2}\times\mathbf{2}\times\mathbf{2}=\left(\mathbf{1}_{A}+\mathbf{3}_{S}\right)\times\left(\mathbf{1}_{A}+\mathbf{3}_{S}\right)=\mathbf{1}_{A}+\mathbf{1}_{S}+\cdots$.
However, this last statement is not correct: the two $SU(2)$ singlets
in the product of four doublets are neither completely symmetric under
permutations nor completely anti-symmetric. Instead, they form an
irreducible doublet of the $S_{4}$ group, which is the relevant permutation
group. In practical terms, this means that we could write the $D_{i}D_{j}D_{k}D_{l}$
quartic interactions as 
\begin{equation}
V^{(4)}=\lambda_{ijkl}^{(1)}\left(D_{i}D_{j}D_{k}D_{l}\right)_{(1)}+\lambda_{ijkl}^{(2)}\left(D_{i}D_{j}D_{k}D_{l}\right)_{(2)}\,,
\end{equation}
where $\left(\cdots\right)_{(1),(2)}$ are the two different contractions
of four doublets and, under a $\pi\in S_{4}$ permutation of the flavour
indices,
\begin{equation}
\left(\begin{array}{c}
\lambda_{\pi\left(ijkl\right)}^{(1)}\\
\lambda_{\pi\left(ijkl\right)}^{(2)}
\end{array}\right)=R\left(\pi\right)\left(\begin{array}{c}
\lambda_{ijkl}^{(1)}\\
\lambda_{ijkl}^{(2)}
\end{array}\right)\,,
\end{equation}
where crucially the twenty-four $R\left(\pi\right)$ matrices cannot
be simultaneously diagonalized. In other words, these matrices form
a 2-dimensional irreducible representation of $S_{4}$. Knowing this,
how many independent couplings are there between $\lambda_{ijkl}^{(1)}$
and $\lambda_{ijkl}^{(2)}$? The answer is $\frac{1}{12}m^{2}(m-1)(m+1)$
and \texttt{Sym2Int} computes this expression for the user:\footnote{For this purpose, \texttt{Sym2Int} uses code from \texttt{Susyno}
which in turn is based on the plethysm function described in the manual
of the \texttt{LiE} package \cite{LiE}.}

\begin{minted}[mathescape,escapeinside=||,bgcolor=lightGray]{mathematica}
gaugeGroup[modelX] ^= {SU2};
fields[modelX] ^= {{"D", {2}, "S", "R", m}};
GenerateListOfCouplings[modelX];
\end{minted}

\noindent produces table \ref{fig:modelX}. Note that in the second
row there is neither an ``\texttt{S}'' nor an ``\texttt{A}'' symmetry:
instead, there is a \texttt{\{2,2\}}. The reason is that irreducible
representations of the $S_{n}$ group can be labeled/associated with
the partitions of the integer $n$. So, for $S_{4}$ the irreducible
representations are \{4\}, \{3,1\}, \{2,2\}, \{2,1,1\}, \{1,1,1,1\}
with dimensions 1, 3, 2, 3 and 1.\footnote{The function \texttt{SnIrrepDim} from \texttt{Susyno} can be used
to find these dimensions \cite{Susyno_GT_Web}.} The program prints for the user the relevant partitions, except in
the cases of \{$n$\} and \{1,1, ...,1\} which are associated to the
completely symmetry (``\texttt{S}'') and the completely anti-symmetric
(``\texttt{A}'') representations. Finally, note that while there
are two independent contractions of four doublets, in the sixth column
of row \#2 of table \ref{fig:modelX}, the last entry is a \texttt{1}.
In reality, this number stands for the number of gauge and Lorentz
contractions with a given permutation symmetry. In the present case,
since \{2,2\} is a 2-dimensional representation of $S_{4}$, there
are $\left(\textrm{last entry =}1\right)\times\left(\textrm{dimension of \{2,2\} =}2\right)$
independent contractions of the gauge (and Lorentz) indices.

\begin{center}
\begin{table}[tbph]
\begin{centering}
\includegraphics[scale=0.82]{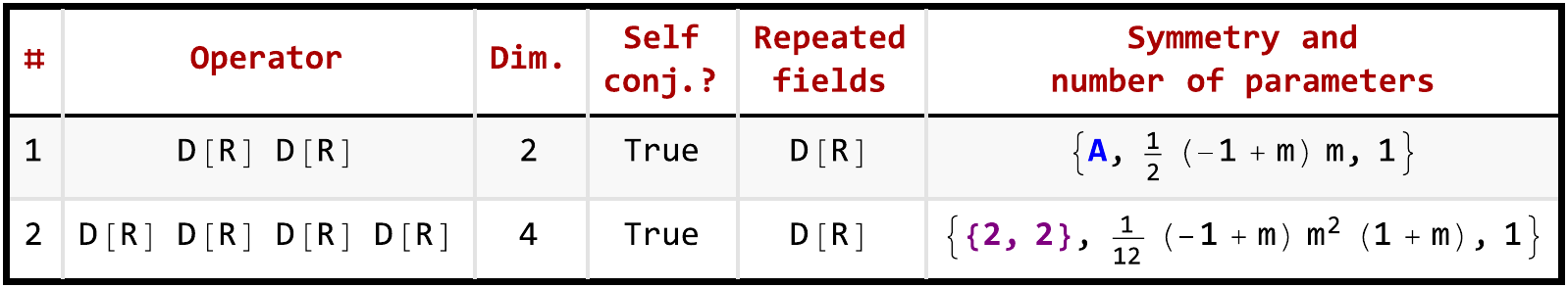}
\par\end{centering}

\protect\caption{\label{fig:modelX}List of interactions in a model with $m$ real
scalar doublets $D$.}
\end{table}

\par\end{center}

\section{Other groups, representations, and program options}

The program is meant to work with any group and gauge/Lorentz representations
(obviously, practical constraints associated to computational time
and/or memory space do apply). To indicate a gauge group the user
only has to write down its name (\texttt{SU2}, \texttt{SU5}, \texttt{SO10},
\texttt{E6}, ...). As for gauge representations of dimension $D$
of some group, in the case of $\boldsymbol{D}$ and $\overline{\boldsymbol{D}}$,
it is enough to provide the numbers $D$ and $-D$ (respectively).
However, there might exist more than one representation with size
$D$, and so primes are added to their names ($\boldsymbol{D'}$,
$\boldsymbol{D''}$, ...). For those representations it is necessary
to use the notation of Dynkin coefficients (see \cite{Susyno_GT_Web, Fonseca:2011sy}
for details). For example, one can indicate the representations $\mathbf{3}$,
$\overline{\mathbf{3}}$, $\mathbf{15}$ and $\overline{\mathbf{15}}$
of $SU(3)$ either with the numbers \texttt{3}, \texttt{-3}, \texttt{15},
\texttt{-15} or with the Dynkin coefficients \texttt{\{1,0\}}, \texttt{\{0,1\}},
\texttt{\{2,1\}}, \texttt{\{1,2\}}, while the $\boldsymbol{15'}$
can only be specified by its Dynkin coefficients, which are \texttt{\{4,0\}}. 

As for Lorentz representations, in most cases fields are either scalars
(\texttt{``S''}), left/right-handed Weyl spinors (\texttt{``L''}/\texttt{``R''})
or perhaps vectors (\texttt{``V''}), in which case it is enough
to provide one of these four strings as input. More generally, irreducible
representations of the Lorentz can be labeled with with two spins
(i.e., two non-negative half-integers) $j_{L}$ and $j_{R}$. \texttt{Sym2Int}
will accept such an input if it is really necessary: for instance
\texttt{``S''}, \texttt{``L''}, \texttt{``R''}, \texttt{``V''}
is equivalent to \texttt{\{0,0\}}, \texttt{\{1/2,0\}}, \texttt{\{0,1/2\}},
\texttt{\{1/2,1/2\}}.

Furthermore, the function \texttt{GenerateListOfCouplings} has several
options.
\begin{itemize}
\item There is support for abelian discrete symmetries through the option\texttt{
DiscreteSym}. Multiplicative conservation of the field charges will
then be enforced by the program: \\
\begin{minted}[mathescape,escapeinside=||,bgcolor=lightGray]{mathematica}
GenerateListOfCouplings[<model>, DiscreteSym->{<charge of field |#1|>, 
|<|charge of field |#2||>|, |...|}];
\end{minted}
\item The function \texttt{GenerateListOfCouplings} not only prints the
list of interactions but it also returns this data in a particular
format (see section \ref{sec:6}). If the table with a model's interactions
is not necessary/desired, it can be suppress with the \texttt{Verbose->False}
option.
\item Hermitian conjugated operators are not shown by default. This can
be changed with the option \texttt{HCTerms->True}.
\item Computing the permutation symmetries which were discussed in section
\ref{sec:4} is time consuming hence, if this information is not needed,
one can use the option \texttt{CalculateSnSymmetries->False} to speed
up the calculation of the list of interactions.
\item As mentioned previously, beyond listing the allowed interactions,\texttt{
Sym2Int} can also calculate how the field components are contracted
in each case. To do that, the option \texttt{CalculateInvariants ->
True} must be used. Note that, depending on the number and size of
the fields, this can be very time consuming. For more details on this
part of the program, the user is referred to the online documentation
\cite{Sym2Int_Web}.
\end{itemize}

\section{\label{sec:6}Saving the program's results for further processing}

The program prints by default a human-readable table with all the
allowed operators. However, one might want to further process \texttt{Sym2Int}'s
results in Mathematica, and in that case one should save and use instead
the output returned by the function \texttt{GenerateListOfCouplings}.
This output consists of a list with the following 8 pieces of information
for each operator:
\begin{enumerate}
\item The number associated to the operator.
\item The combination of fields which are interacting. Each field is identified
by its position in the list provided by the user as input (if the
field does not appear conjugated in the operator) or minus its position
in the list provided by the user as input (if the field does appear
conjugated in the operator).
\item The mass dimension of the operator.
\item Is the combination of fields self-conjugated?
\item The list of fields which appear more than once in the operator (the
identification of each field is done as in 2.).
\item Permutation symmetry and number of parameters associated to the operator.
\item Information on the explicit expression(s) of the gauge and Lorentz
invariant contraction (or contractions if there are several) of the
fields which appear in the operator. In other words, it provides the
information needed to write down the Lagrangian. (Requires the use
of the option \texttt{CalculateInvariants -> True}; otherwise this
slot in the list will contain the `\texttt{Null}' value.)
\item The exact same data/format/style as the relevant row of the table
printed by the function \texttt{}~\\
\texttt{GenerateListOfCouplings}.
\end{enumerate}
To get a quick view of what this means in practice, consider the Standard
Model as defined previously in section \ref{sec:3}. The code

\begin{minted}[mathescape,escapeinside=||,bgcolor=lightGray]{mathematica}
operatorsSM=GenerateListOfCouplings[SM];
Grid[operatorsSM, Frame -> All]
\end{minted}

\noindent yields the result 

\begin{center}
\includegraphics[scale=0.82]{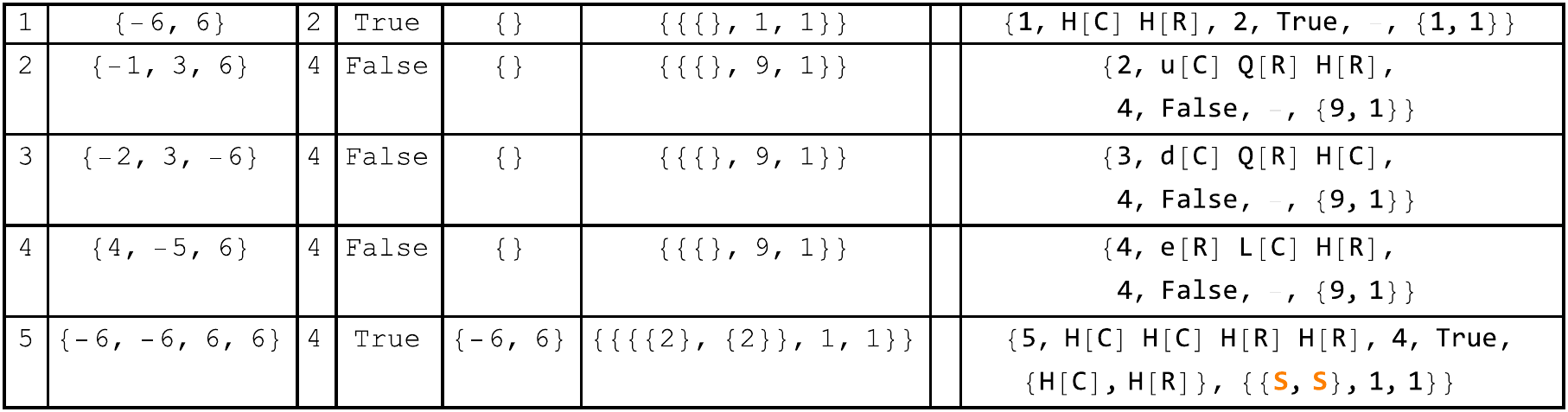}\,.
\par\end{center}

\section*{Acknowledgements}

I would like to thank the organizers of the DISCRETE 2016 Symposium
in Warsaw, where this computer code was first presented. I am also
grateful to the other participants which provided interesting suggestions
concerning the program. Lastly, this work was financially supported
by the Spanish state through the grants Juan de la Cierva-formación
FJCI-2014-21651, FPA2014-58183-P, Multidark CSD2009-00064 and SEV-2014-0398
(from the \textit{Ministerio de Economía, Industria y Competitividad}),
as well as the grant PROMETEOII/2014/084 (from the \textit{Generalitat
Valenciana}).

\end{document}